\documentstyle[psfig]{caosp}
\begin{document}
%
%
%%%%%%%%%%%%%%%%%%%%%%%%%%%%%%%%%%%%%%%%%%%%%%%%%%%%%%%%%%%%%%%%%%%%%%%%%%%%%%
%                       E D I T O R I A L   N O T E S                        %
% Next 3 lines are not obligatory, they will be inserted by the editors      %
%%%%%%%%%%%%%%%%%%%%%%%%%%%%%%%%%%%%%%%%%%%%%%%%%%%%%%%%%%%%%%%%%%%%%%%%%%%%%%
%pubyear{1993}
%volume{23}
%firstpage{7}
%%%%%%%%%%%%%%%%%%%%%%%%%%%%%%%%%%%%%%%%%%%%%%%%%%%%%%%%%%%%%%%%%%%%%%%%%%%%%%
%              R U N N I N G   P A G E   H E A D I N G S                     %
% The odd pages headings (except that of the title page) are produced        %
% automatically and contain the title. If, however, the title of your article%
% is too long, the running head will be ommitted in the printout; you can    %
% then make your own running title by using the \htitle command putting the  %
% shortened title between the curly brackets; should you want to             %
% see how it works remove simply the % sign from the beginnig of that line.  %
% To produce the even pages headings use the \hauthor command as shown below.%
%%%%%%%%%%%%%%%%%%%%%%%%%%%%%%%%%%%%%%%%%%%%%%%%%%%%%%%%%%%%%%%%%%%%%%%%%%%%%%
\htitle{High-Energy Phenomena in Magnetic CP Stars}
\hauthor{Stephen A. Drake}
%%%%%%%%%%%%%%%%%%%%%%%%%%%%%%%%%%%%%%%%%%%%%%%%%%%%%%%%%%%%%%%%%%%%%%%%%%%%%%
%                       T I T L E                                            %
% You should  not use just capital letters in the title. Don`t end the       %
% title by "." (dot)                                                         %
%%%%%%%%%%%%%%%%%%%%%%%%%%%%%%%%%%%%%%%%%%%%%%%%%%%%%%%%%%%%%%%%%%%%%%%%%%%%%%
\title{High-Energy Phenomena in Magnetic CP Stars as Revealed by
their X-Ray and Radio Emission}
%%%%%%%%%%%%%%%%%%%%%%%%%%%%%%%%%%%%%%%%%%%%%%%%%%%%%%%%%%%%%%%%%%%%%%%%%%%%%%
%                       S U B T I T L E                                      %
% You can use subtitle, with command \subtitle similiar to \title command    %
%%%%%%%%%%%%%%%%%%%%%%%%%%%%%%%%%%%%%%%%%%%%%%%%%%%%%%%%%%%%%%%%%%%%%%%%%%%%%%
%%%%%%%%%%%%%%%%%%%%%%%%%%%%%%%%%%%%%%%%%%%%%%%%%%%%%%%%%%%%%%%%%%%%%%%%%%%%%%
%                   A U T H O R  N A M E S                                   %
% Authors` names are separated by the \and commmand their institutes are     %
% asigned by the \inst{n} command                                            %
%%%%%%%%%%%%%%%%%%%%%%%%%%%%%%%%%%%%%%%%%%%%%%%%%%%%%%%%%%%%%%%%%%%%%%%%%%%%%%
\author{Stephen A. Drake \inst{1}}
%%%%%%%%%%%%%%%%%%%%%%%%%%%%%%%%%%%%%%%%%%%%%%%%%%%%%%%%%%%%%%%%%%%%%%%%%%%%%%
%           I N S T I T U T E S'  A D D R E S S E S                          %
% The affiliation of authors is generated by the \institute command, the \and%
% command being again used to separate individual addresses.                 %
% The following commands may be used for the three institutes in question:   %
%               \lomnica        for      AsU SAV, Tatranska Lomnica          %
%               \blava          for      AsU SAV, Bratislava                 %
%               \ondrejov       for      AsU CAV, Ondrejov                   %
%%%%%%%%%%%%%%%%%%%%%%%%%%%%%%%%%%%%%%%%%%%%%%%%%%%%%%%%%%%%%%%%%%%%%%%%%%%%%%
\institute{HEASARC, Goddard Space Flight Center\\Greenbelt, MD, U.S.A.}

%%%%%%%%%%%%%%%%%%%%%%%%%%%%%%%%%%%%%%%%%%%%%%%%%%%%%%%%%%%%%%%%%%%%%%%%%%%%%%
%                        D A T E                                             %
% Date inserted here will be the date of your paper receiving                %
% or you can use \today command instead explicit date in brackets            %
%%%%%%%%%%%%%%%%%%%%%%%%%%%%%%%%%%%%%%%%%%%%%%%%%%%%%%%%%%%%%%%%%%%%%%%%%%%%%%
\date{\today}
\maketitle
%%%%%%%%%%%%%%%%%%%%%%%%%%%%%%%%%%%%%%%%%%%%%%%%%%%%%%%%%%%%%%%%%%%%%%%%%%%%%%
%                        A B S T R A C T,  K E Y W O R D S                   %
% Here is shown how to write the abstract                                    %
% Write your keywords using command \keywords  the thesaurus from Astron.    %
% Astrophys. Abstracts.                                                      %
%%%%%%%%%%%%%%%%%%%%%%%%%%%%%%%%%%%%%%%%%%%%%%%%%%%%%%%%%%%%%%%%%%%%%%%%%%%%%%
\begin{abstract}
Before 1985, attempts to detect radio or X-ray emission from Magnetic CP 
(MCP) stars were either fruitless or ambiguous. However, more successful
results have been obtained in the last dozen years: (i) Radio emission has
now been detected from $\sim 35$ MCP stars of the Helium-peculiar and
Silicon-strong subclasses, with a functional dependence of radio luminosity
$L_{R} \propto T_{eff}^7 H_s^{1.2} P_{rot}^{-0.6}$,
where $T_{eff}$ is the effective temperature, $H_s$ is the surface magnetic 
field strength, and $P_{rot}$ is the rotational period; rotational modulation of
the radio emission has also been observed for several MCP stars. All of this 
evidence suggests that it is the MCP stars themselves, not close companions,
that are responsible for the radio emission; (ii) The X-ray emission properties
of MCP stars are however still poorly characterized: although a moderate
number ($\sim 18$) have by now been associated with X-ray sources,
the lack of a clear correlation of this X-ray emission with other stellar
parameters has made it difficult to rule out the close companion hypothesis.
\end{abstract}
%%%%%%%%%%%%%%%%%%%%%%%%%%%%%%%%%%%%%%%%%%%%%%%%%%%%%%%%%%%%%%%%%%%%%%%%%%%%%%
%                       S E C T I O N I N G                                  %
% A section starts with the command \section as shown below, with the title  %
% in Initial Capitals and lowercase only. Do not number the sections - let   %
% LaTeX do that for you - and do not end them by ".".                        %
%%%%%%%%%%%%%%%%%%%%%%%%%%%%%%%%%%%%%%%%%%%%%%%%%%%%%%%%%%%%%%%%%%%%%%%%%%%%%%
\section{Introduction}
%%%%%%%%%%%%%%%%%%%%%%%%%%%%%%%%%%%%%%%%%%%%%%%%%%%%%%%%%%%%%%%%%%%%%%%%%%%%%%
%                       L A B E L                                            %
% Label command is very convenient for you when referring to secctions,      %
% subsections,..., tables, figures as well as equations in your article (see %
% commands \ref and \pageref). In case figure and table environments         %
% the \label command should always be put after the \caption command to      %
% preserve a proper numbering. When using the \label command you must compile%
% the file twice to get proper cross-references.                             %
%%%%%%%%%%%%%%%%%%%%%%%%%%%%%%%%%%%%%%%%%%%%%%%%%%%%%%%%%%%%%%%%%%%%%%%%%%%%%%
\label{intr}
%%%%%%%%%%%%%%%%%%%%%%%%%%%%%%%%%%%%%%%%%%%%%%%%%%%%%%%%%%%%%%%%%%%%%%%%%%%%%%
%                       P A R A G R A P H                                    %
% To generate a paragraph just leave a blank line after the last sentence    %
% of the preceding paragraph as shown below.                                 %
%%%%%%%%%%%%%%%%%%%%%%%%%%%%%%%%%%%%%%%%%%%%%%%%%%%%%%%%%%%%%%%%%%%%%%%%%%%%%%
Magnetic chemically peculiar stars (hereafter MCP stars) have been
recognized as a distinct subclass of chemically peculiar (CP) stars since
the 1960s. The strong (kiloGauss) and predominantly dipolar magnetic fields 
of these stars have been considered to be possible sites of
high-energy phenomena, both thermal and non-thermal, for almost as long.
The closed field regions of MCP stars can, due to the high magnetic pressure,
clearly confine any ionized plasma that flows from the stellar surfaces, and 
interaction between such putative stellar winds and the strong magnetic fields
might also result in the acceleration of ions to extremely large energies.
An important theoretical paper which clearly presented much of this physics
was published by Havnes \& Goertz in 1984 (q.v.). It was quickly realized that,
since the stars with the higher effective temperatures have the strongest
radiation-driven winds, the B-type MCP stars (i.e., the helium-peculiar
and silicon-strong subclasses) were much more likely to
exhibit high-energy phenomena than the `classical' A-type MCP stars (i.e.,
the CrSrEu-strong subclass).
Shore in a series of papers (Shore 1987; Shore {\it et al.} 1987; Shore
\& Brown 1990) followed up on earlier work to propose that
the He-pec stars had predominantly polar stellar winds as well as 
magnetospherically trapped plasma in the equatorial regions, and used this 
model to explain the variation with rotational phase of ultraviolet emission
lines formed at temperatures $T_e \sim 10^5$K. Earlier papers (e.g., Groote 
\& Hunger 1982) had fit the hydrogen Balmer lines of the He-strong star
$\sigma$ Ori E with a model in which plasma with $T_e \sim 10^4$K was
confined in clouds in the equatorial regions.

In the present review, I will restrict my discussion to observations of
MCP stars aimed at detecting plasma at temperatures $\ga 10^6$K
using X-ray telescopes, and to observations to detect high-energy, thermal
or nonthermal electrons
using radio telescopes. Thus, these studies can be regarded as
complementing the aforementioned optical and UV studies that have helped to 
constrain the characteristics of the cooler components of the circumstellar
plasma. I will briefly review the early studies in the radio and X-ray
energy ranges and then discuss what has been newly learned in the last decade.

\section{Radio Continuum Observations of MCP Stars}
\label{Radio}
The first radio studies dedicated to MCP stars were those of Trasco {\it
et al.} (1970) and Kodaira \& Fomalont (1970). These and subsequent radio
surveys in the 1970s were generally made using single-dish radio telescopes
operating at cm wavelengths, and had spatial resolutions and positional
accuracies of a few arcmin, and detection limits of 10 - 50 mJy,
equivalent to radio luminosities $L_R$ of $10^{17-18}$ erg s$^{-1}$ Hz$^{-1}$
at a distance of 100 pc. Trasco {\it et al.} failed to detect 3 MCP stars,
including the well-studied Ap star $\alpha^2$ CVn, while Kodaira
\& Fomalont also had no convincing detections in a sample of 12 MCP stars,
although they did report a 50 mJy ($2\sigma$) excess at the position of the MCP
star with the largest measured magnetic field, i.e., Babcock's Star or
HD 215441. A few years later, Altenhoff {\it et al.} (1976) reported a 
further 13 non-detections of MCP stars in the radio band. These negative
results, indicating that strong radio emission is not present in MCP stars, 
dampened interest in this field for almost a decade.

Radio astronomy was revolutionized in the late 1970s and 1980s
by the start of operations of
radio interferometers such as the Very Large Array (VLA) and the Australia
Telecope Compact Array which have orders of magnitude improvement
over single-dish telescopes in their sensitivity and angular resolution:
observations with arcsecond accuracy and sub-mJy sensitivity became routine.
For example, the VLA can detect a source of 0.1 mJy ($L_R \sim 10^{15}$ erg 
s$^{-1}$ Hz$^{-1}$ at 100 pc) at 3.6 or 6 cm in less than an hour, meaning
that a survey 100 times deeper than the 1970s era programs could be done
in less than a day. Also, measuring
circularly polarized radio signals with the VLA is quite straightforward. 
Somewhat later, improvements in Very Long Baseline Interferometry 
(VLBI) techniques that culminated in the construction of the Very Long 
Baseline Array made radio observations with milliarcsecond spatial accuracy 
possible.

Somewhat embarrassingly, the first reliable radio detection of an MCP star
was serendipitous: in the field of the `normal' O9.5 V star $\sigma$
Ori A, Dave Abbott and his collaborators detected a 3 mJy radio source that was
spatially coincident with its visual companion, $\sigma$ Ori E, a He-strong
(HeS) MCP star. Stimulated by this discovery, a group of us surveyed 14 MCP
stars, confirming that $\sigma$ Ori E was a flat-spectrum radio source,
and finding another MCP radio source, HR 1890 = HD 37017 (Drake {\it et al.} 
1985). Both detected
stars were HeS stars in the Orion OB1 Association that had measured kiloGauss
(kG) photospheric magnetic fields, and had an implied $\L_R \sim 6 
\times 10^{17}$
erg s$^{-1}$ Hz$^{-1}$. The flat spectral index of $\sigma$ Ori E between
2 and 6 cm implied that the emission was either optically thin free-free
(ff) or nonthermal in nature. An expansion of this VLA program to 34 MCP
stars was reported by Drake {\it et al.} (1987) where 3 more radio detected 
MCP stars were listed, including a third HeS star, and two much later spectral
type Si-type MCP stars, HD 215441 and IQ Aur. None of the later-type SrCrEu
MCP stars were detected, however, with upper limits as low as $3 \times 
10^{14}$ erg s$^{-1}$ Hz$^{-1}$, i.e., 3 dex lower than
the radio luminosities of the HeS stars, in either the above study or a later
one of Willson {\it et al.} (1988). 

The 3 HeS stars appeared to form a rather homogeneous sample of 
high-luminosity radio emitters, and detailed studies
of these stars showed that their radio fluxes were varying on a timescale of
hours. This last property essentially clinched that the emission was nonthermal
in nature, although another property of some types of nonthermal emission,
viz. detectable levels of circular polarization, was not confirmed to be
present in these radio sources. The other 2 radio-detected later-type stars 
were much weaker
radio emitters ($\L_R \sim 10^{16-17}$ erg s$^{-1}$ Hz$^{-1}$) than the HeS
stars and it was not clear whether their emission mechanisms were similar.
Drake {\it et al.} 1987 proposed a schematic `radiation-belt' model for the 
HeS radio emitters in which the radio emission was optically thick 
gyrosynchrotron emission of mildly relativistic, nonthermal electrons that
were sited in the closed-field magnetospheric regions at several stellar
radii. This model required essentially continuous injection of nonthermal
particles into the magnetospheric regions with a density profile that
increased with radius so as to fit the observed radio spectra.
Phillips \& Lestrade (1988) confirmed the nonthermal nature of the HeS star
radio emission using VLBI techniques when they failed to resolve the radio
emission from 2 HeS stars, which enabled them to derive lower limits to
the radio brightness temperatures of $10^9$K that were inconsistent with
thermal models.

Meanwhile, Andr\'{e} {\it et al.} (1988, 1991) developed a model to explain an 
optically obscured, radio-emitting object (S1) in the $\rho$ Oph star-forming
region in which they
suggested that it was a young MCP star. (Notice, however, that there is no 
other direct evidence that $\rho$ Oph S1 is actually an MCP star).
In this model, which would seem to be also applicable to the previously 
detected HeS star radio emitters, the radio emission was due to 
optically thin gyrosynchrotron emission from a torus of about 10 stellar
radii, while the source of the nonthermal electrons was 
the magnetotail or current sheet formed where the predominantly polar
stellar wind of this star terminates the closed-field magnetospheric regions.
Andr\'{e} {\it et al.} also argued that plasma in the inner magnetosphere
would be heated to coronal temperatures, thus explaining the observed high
($10^{31-32}$ erg s$^{-1}$) X-ray luminosity of this object. 
In their second paper, this group reported VLBI observations of S1
which implied a radio-emission region of $\sim 6$ stellar radii and a
brightness temperature of $10^8$K, and adjusted their model slightly to fit
these new constraints. 

Leone (1991) showed that the radio emission from $\sigma$ Ori E and HR 1890
varies with the phase of the magnetic field (i.e., the rotational phase), 
and proposed a model in which
the radio emission was due to gyroresonance emission from low-temperature
($T_e \ll 10^6$K) electrons in the polar wind regions. This model also
predicted that the radio emission should exhibit zero circular polarization.
Leone \& Umana (1993) obtained more radio data on these two stars,
strengthening the observed correlation with magnetic phase, and made a
comparison of the rival models for MCP star radio emission.

Linsky {\it et al.} (1992) updated the Drake {\it et al.} (1987) VLA survey
results, reporting detection of 3 out of 9 HeS stars (all in Orion OB1) with
$\log L_R  \sim 16.8 - 17.9$, 13 out of 38 HeW/Si stars (9 in Sco OB2) with
$\log L_R  \sim 14.7 - 17.2$, and 0 out of 14 cool Ap stars with 
$\log L_R < 14.7$. In addition, in 4 of the detected stars variable
circular polarization at levels of up to $30\%$ was detected, ruling out (at
least in these cases) the gyroresonance emission model. Using both the
radio detections and the significant non-detections, Linsky {\it et al.}
found that the radio emission obeyed a global relation $L_R \propto
\dot{M}^{0.4} B^{1.1} P_{rot}^{-0.3}$, where the mass-loss rate $\dot{M}$ was
derived from the stellar effective temperature $T_{eff}$ using a functional
dependence derived by Lamers for normal (i.e., nonmagnetic) hot stars.
They revised the Drake {\it et al.} (1987) model to one in which the emission 
(still attributed to optically thick gyrosynchrotron emission) originated in 
two tori above and below the magnetic equator, the radii of which increased
with increasing wavelength from 3 stellar radii at 6 cm to $\sim 10$ stellar
radii at 20 cm. They agreed with Andr\'{e} {\it et al.} that the current
sheet is the likely source of the nonthermal electrons, and, under this
assumption, estimated the total magnetic energy content of the current sheet
as $\propto \dot{M}^{0.7} B^{0.7} P_{rot}^{-1.3}$, i.e., rather similar
to the functional dependence of the observed radio emission. 

Leone {\it et al.} (1994) detected 3 more Si-type Bp stars as radio emitters
and presented more evidence that the radio maxima for the well-observed 
systems occured at magnetic field extrema. Later, Leone {\it et al.} (1996)
discussed the 1.3 to 20 cm radio spectra of 7 MCP stars, finding rather
diverse spectral properties for these stars that imply that in some (but not
all) MCP stars the radio emission is absent or suppressed in the inner
magnetospheric regions. They also reported non-detections at 1.3 mm
wavelength of 3 of the MCP stars.

Recently, Drake {\it et al.} (1998a) have detected 16 more HeW/Si star
radio emitters, bringing the total number of known radio-emitting MCP stars
to about 35, i.e., this is a well-established class of radio-emitting
stars. Using a subset of 38 MCP stars for which well-defined fundamental
parameters are available, 24 of which were detected as radio sources, and
14 of which were not, Drake {\it et al.} found an empirical relation that is 
accurate to $\sim \pm 0.6$ dex:
\begin{eqnarray}
\log(L_R) & \approx & 15.1 + (6.85 \times \log(T_{eff}/10^4)) + (1.20 \times
\log(H_s/\rm{kG})) \nonumber \\
         &         & - (0.60 \times \log(P_{rot}/\rm{days}),
\end{eqnarray}
where $H_s$ is the effective surface magnetic field.

\section{X-Ray Observations of MCP Stars}
\label{Xray}
MCP stars have been proposed as possible counterparts to X-ray sources for
at least 20 years (e.g., Groote {\it et al.} 1978, Cash {\it et al.} 
1979), but it has proven
unexpectedly hard to confirm them as a `real' class of X-ray sources. 
This is partly because X-ray emission in stars is quite ubiquitous (with
the possible exception of A-type stars), and, thus contaminating emission
from low-mass binary companions is always a possible factor. Nevertheless,
there were a handful of candidate MCP X-ray sources that were identified
in {\it Einstein} and {\it EXOSAT} X-ray observations in the 1980s, e.g.,
Cash \& Snow (1982), Cutispoto {\it et al.} (1990), but no obvious
pattern of MCP star fundamental properties versus X-ray properties was
evident. The MCP stars which might have been predicted to have the best
chance of having intrinsic X-ray emission, viz. the HeS stars, do in 
fact appear to be X-ray emitters (cf. Drake {\it et al.} 1987),
but only at levels that are similar to those seen in non-magnetic stars of
the same spectral types, i.e., the HeS star X-ray emission appears to be
related to the stellar wind region, and not to the magnetosphere. Thus,
by 1990, MCP stars had still not been established as a class of intrinsic
X-ray emitters, although there were some interesting `suspects', e.g., 
some of the 12 B6-A3 stars in Ori OB1 that were detected as X-ray sources
by Caillault \& Zoonematkermani (1989) may be MCP in nature.
Given that the X-ray emission properties of normal B3-A9 stars were not
well-known at this time (except that their X-ray luminosities were too low
to detect typically), better observations were clearly required.

The launch of the {\it ROSAT} X-ray observatory in 1990 revolutionized X-ray
astronomy, primarily due to its being 10-100 times more sensitive than
earlier X-ray observatories: now, stars with $\log L_X \geq 28.0$ only
slightly stronger than that of the Sun ($\log L_X \sim  27.3 \pm 0.5$)
could easily be detected at distances of up to 100 pc. This enormously
increased the number of MCP stars which could be searched for X-ray
emission. The subsequent launch in 1993 of the {\it ASCA} X-ray observatory
enabled, for the first time, moderate-resolution X-ray spectroscopy
($E/\Delta E \sim 30$ compared to $\sim 3$ for {\it ROSAT}) to be performed on 
such weak X-ray sources as MCP stars appear to be.

A theoretical paper which appeared around this time (Usov \& Melrose 1992)
proposed that the X-ray emission observed in luminous (nonmagnetic) 
OB stars was in fact due to a current sheet formed between the known strong
stellar winds of these stars and hypothetical small magnetospheres (with
implied photospheric field strengths of $\sim 100$ Gauss lying below present
detection limits). The predicted functional dependence of X-ray emission
on fundamental properties for this mechanism, viz., $ L_X \propto \dot{M}^{9/4}
V_{wind}^{-3/4} H_s^{-1/2} R_{\ast}^{-3/2}$,
means that, extrapolating this model to the relatively low mass flux and
high magnetic field MCP stars, X-ray luminosities at undetectable levels
($\log L_X \sim 23 - 24$) are predicted. However, it should be noted that (a)
this model does not appear to be widely accepted as applying to nonmagnetic
hot stars, and (b) its extrapolation to MCP stars may be rather risky.\\
\indent Despite this caveat, Drake {\it et al.} (1994) studied the {\it ROSAT}
All-Sky Survey X-ray database at the positions of $\sim 100$ radio-observed
MCP stars, with a detection level of $\log L_x \sim 29.4$ at 100 pc (note
that this is significantly poorer than that achievable in pointed {\it ROSAT}
observations). After eliminating a handful of cases in which binary companions
to MCP stars seemed the more likely candidates for being the X-ray emitters,
only 6 out of 100 MCP stars (2 of which are HeS stars and the remainder are
HeW/Si stars) seemed to be candidates for being intrinsic X-ray
emitters, a very low detection rate not inconsistent with a true rate of $0\%$.
They concluded that MCP stars are clearly not a class of X-ray emitters at
levels in excess of $\log L_X \sim 30$, and furthermore found no evidence
that there was any correspondence between radio and X-ray emission in these 
stars, in that many radio-bright stars were not detected as X-ray sources
(and {\it vice versa}). A similarly low detection rate was reported by Leone 
(1994) based on a re-examination of the archival {\it Einstein} data on MCP
stars: he found only 4 out of 90 MCP Bp stars were detected as X-ray sources,
but noted that 3 of these 4 stars were actually binary systems.\\
\indent Subsequent to this, only a few other papers on this subject have 
appeared: perhaps the most interesting being (a) the {\it ROSAT}
studies of the young ($10^8$ years) open cluster NGC 2516 by Dachs \&
Hummel (1996) and Jeffries {\it et al.} (1997) in which 4-6 of the known
MCP stars in this cluster were detected as X-ray sources (the second paper
concluded that this emission was probably intrinsic, although binary 
companions cannot be entirely ruled out), and (b) the discovery by Gagn\'{e}
{\it et al.} (1997) of X-ray emission from the Orion Trapezium star
$\theta^1$ Ori C that varies periodically on a 15 day timescale: these
authors suggested that this O7 V star could be a high-mass example of
an MCP star (but this star has not, of course, been otherwise proven to be
an MCP star). In some of my unpublished work (Drake {\it et al.} 1998b),
I have extended the program of Drake {\it et al.} (1994) to include archival
pointed {\it ROSAT} observations, and have about 18 candidates among MCP
stars for intrinsic X-ray emission, including Babcock's Star (with $\log
L_X \sim 29.2$) and 13 other HeW/Si stars, and 4 SrCrEu-type stars. Despite
the increase in this sample size, there is still little evidence that the
X-ray emission of these stars correlates with any other known property,
meaning that it is still impossible to predict which MCP stars will be
X-ray emitters {\it a priori}. The logarithmic ratio of the 
X-ray to the radio emission varies
from $<11.9$ to $14.5$, for example, which should be compared to the value
$\log (L_X/L_R) \sim 15.0 \pm 0.5$ found for active late-type stars with similar
radio luminosities to those of the MCP stars. Thus, all MCP stars appear
to be X-ray weak relative to their radio emission strength compared to the
late-type stars: this is not a trivial comparison, because it should be noted
that the mechanism for the production of the radio emission in both of
these classes of stars, i.e., optically thick gyrosynchrotron emission from
mildly relativistic electrons, is essentially the same, although the geometry
is certainly different. 

In Drake {\it et al.} (1998b), X-ray spectra
of some of these candidate X-ray emitting MCP stars are also discussed.
These spectra look, in general, very similar morphologically to the X-ray 
spectra of other classes of stars,
i.e., there is nothing unusual in MCP star X-ray spectral shapes that could
be used as a discriminator. One exception is the {\it ROSAT} spectrum of
the Si star HR 5624, which appears to be best fit with a Si abundance in the 
X-ray emitting plasma of $\sim 15$ times the solar photospheric value and
which is similar to the levels of overabundance of Si in the photospheres
of Si stars. In a follow-up {\it ASCA} observation of this same star
obtained a couple of years later, I derived a much lower Si overabundance
(about twice solar): notice, however, that due to the lower sensitivity of
{\it ASCA} compared to {\it ROSAT}, this second observation was essentially
an average over an entire rotational phase of this rapidly rotating star
($P_{rot} = 0.88$ days), whereas the {\it ROSAT} observation was made over
only a small fraction of the phase. Thus, the possibility that the X-ray
emitting plasma may be overabundant in Si in a time-variable way cannot be
ruled out.

\section{Conclusions}
\label{Conc}
There is an essential difference between the radio and the X-ray properties
of the MCP stars, with the former being well-established and the latter
quite the opposite. Nonthermal radio emission is prevalent among the B-type MCP
stars, and the basic emission mechanism is almost certainly gyrosynchrotron
emission from mildly relativistic nonthermal electrons.
Their radio emission properties are predictable from their basic properties 
(cf. eq. 1), and, in general terms, explainable
using current models, although detailed models which fit the radio spectra
and circular polarizarion as a function of aspect for specific stars are
still pending. 
Their X-ray emission properties are still uncertain: the existence
of significant amounts of plasma at $\ga 10^6$K has still not
been confirmed to be a common feature in MCP stars despite 20 years of
observations. The presence of X-ray emission is not predictable from either
their MCP type or their radio properties. In this sense, their status as a 
class of intrinsic X-ray emitters is, in my opinion, still not demonstrated. If
the existing candidate X-ray emitters are intrinsic, then a revision of the
Usov \& Melrose (1992) model is clearly required, since, in its present form,
it is underpredicting the X-ray luminosities by 7 or 8 orders of magnitude.
Conclusive evidence for the intrinsic hypothesis might be (a) the discovery
of a distinct X-ray spectral signature, such as elemental abundances showing
a similar pattern to the underlying photospheric abundances, or (b) the
finding of an X-ray modulation with a similar period to that of the magnetic
light-curve, or (c) the acquisition of enough X-ray detections among MCP stars
that a correlation with some other stellar properties may be found.
%%%%%%%%%%%%%%%%%%%%%%%%%%%%%%%%%%%%%%%%%%%%%%%%%%%%%%%%%%%%%%%%%%%%%%%%%%%%%%
%                       A C K N O W L E D G E M E N T S                      %
% Next lines show you how to write acknowledgements                          %
%%%%%%%%%%%%%%%%%%%%%%%%%%%%%%%%%%%%%%%%%%%%%%%%%%%%%%%%%%%%%%%%%%%%%%%%%%%%%%
% You must leave a blank line before the \acknowledgements command!

\acknowledgements
% Do not leave a blank line here! <---------------------->
The author is extremely thankful to Dr. Jeffrey Linsky for their longstanding
collaboration on these topics. This work has been supported in part by
NASA Guest Investigator contracts from the {\it ROSAT} and {\it ASCA}
missions.

%%%%%%%%%%%%%%%%%%%%%%%%%%%%%%%%%%%%%%%%%%%%%%%%%%%%%%%%%%%%%%%%%%%%%%%%%%%%%%
%                       R E F E R E N C E S                                  %
% References should start with the \begin{thebibliography}{} command, leaving%
% the last curly brackets empty.                                             %
%%%%%%%%%%%%%%%%%%%%%%%%%%%%%%%%%%%%%%%%%%%%%%%%%%%%%%%%%%%%%%%%%%%%%%%%%%%%%%

%%%%%%%%%%%%%%%%%%%%%%%%%%%%%%%%%%%%%%%%%%%%%%%%%%%%%%%%%%%%%%%%%%%%%%%%%%%%%%
%                       H A P P Y E N D                                      %
% Your LaTeX source text must be ended by the line:                          %
%%%%%%%%%%%%%%%%%%%%%%%%%%%%%%%%%%%%%%%%%%%%%%%%%%%%%%%%%%%%%%%%%%%%%%%%%%%%%%
\end{document}